\documentclass[aps,preprint,amsmath,amssymb,showpacs]{revtex4}

\usepackage{graphicx}
\begin{document}

\title{Anomalous Higgs couplings in $e\gamma$ collision with initial beam
and final state polarizations}

\author{\.{I}nan\c{c} \c{S}ahin}
\email[]{isahin@science.ankara.edu.tr} \affiliation{Department of
Physics, Faculty of Sciences, Ankara University, 06100 Tandogan,
Ankara, Turkey}

\begin{abstract}
We investigate the constraints on the anomalous {\it WWH} couplings
through the process $e^{-}\gamma \to \nu_{e} W^{-} H$. Considering
incoming beam polarizations and the longitudinal and transverse
polarization states of the final W boson, we find 95\% confidence
level limits on the anomalous coupling parameters with an integrated
luminosity of 500 $fb^{-1}$ and $\sqrt{s}$= 0.5 and 1 TeV energy. We
show that initial beam and final state polarizations highly improve
the sensitivity limits of the anomalous coupling parameters $b_{W}$
and $\beta_{W}$.
\end{abstract}


\maketitle

\section{Introduction}
The Standard model (SM) of particle physics has been proven to be
successful in the energy scale of the present colliders.
Experimental results obtained from recent experiments at CERN LEP
and Fermilab Tevatron confirm the $SU(3)_{C}\times SU(2)_{L}\times
U(1)_{Y}$ gauge structure of SM. However, the Higgs boson which is
crucial for electroweak symmetry breaking and mass generation has
not been observed, and one of the main goals of future experiments
is to pursue its trace. Although the exact value of the Higgs  mass
is unknown there are experimental bounds for it. Direct searches
from ALEPH, DELPHI, L3, and OPAL collaborations at CERN LEP provide
a lower bound on the Higgs mass of $m_{H}>114.4$ GeV \cite{Barate}.
Indirect experimental bounds for the Higgs mass are obtained from
fits to precision measurements of electroweak observables. The
current best fit value sets an upper bound of $m_{H}<186$ GeV
\cite{LEPEWWG}.

If a Higgs boson is detected at future TeV scale colliders, it will
be crucial to test its couplings to the SM particles. Precision
measurements of Higgs couplings may give us some hints for new
physics beyond  the SM. It is argued that CERN LHC has a potential
to search for the Higgs boson in the entire mass range allowed. In
case it is found at CERN LHC, precision measurements of its
couplings to the SM particles can be obtained at future linear
$e^{+}e^{-}$ colliders. After linear $e^{+}e^{-}$ colliders are
constructed its operating modes of $e\gamma$ and $\gamma\gamma$ are
expected to be designed \cite{akerlof}. A real gamma beam is
obtained through Compton backscattering of laser light off the
linear electron beam where most of the photons are produced at the
high energy region. The luminosities for $e\gamma$ and
$\gamma\gamma$ collisions turn out to be of the same order as the
one for $e^{+}e^{-}$ \cite{Ginzburg}, so the cross sections for
photoproduction processes with real photons are considerably larger
than the virtual photon case. Polarizability of the real gamma beam
is an additional advantage for polarized beam experiments. Therefore
$e\gamma$ and $\gamma\gamma$ collisions should be discussed as
complementary to $e^{+}e^{-}$ collisions.

In this work we analyzed the anomalous {\it WWH} vertex in the
$e\gamma$ collision. We consider the process $e^{-}\gamma \to
\nu_{e} W^{-} H$. This process isolates the {\it WWH} vertex and
gives us the opportunity to study the {\it WWH} vertex independent
from {\it ZZH}. On the other hand in $e^{+}e^{-}$ collisions,
observables depending on {\it WWH} also receive contributions from
the {\it ZZH} and it is very difficult to dissociate {\it WWH} from
{\it ZZH} \cite{Choudhury1}.

Anomalous {\it WWH} couplings can be investigated in a model
independent way by means of the effective Lagrangian approach
\cite{Gonzalez-Garcia,Barger, Choudhury1,Choudhury2}. In writing
effective operators we employ the formalism of
\cite{Choudhury1,Choudhury2}. Imposing Lorentz invariance and gauge
invariance and retaining up to dimension six operators in the
effective Lagrangian, the most general coupling structure is
expressed as,

\begin{eqnarray}
\Gamma^{V}_{\mu\nu}=i\widetilde{g}_{V}\left[a_{V}g_{\mu\nu}
+\frac{b_{V}}{m_{V}^{2}}(k_{2\mu}k_{1\nu}-g_{\mu\nu}k_{1}.k_{2})
+\frac{\beta_{V}}{m_{V}^{2}}\epsilon_{\mu\nu\alpha\beta}k_{1}^{\alpha}k_{2}^{\beta}\right]
\end{eqnarray}

with
\begin{eqnarray}
\widetilde{g}_{W}=g_{W}m_{W} \;,\;\;\;\;\;\;
\widetilde{g}_{Z}=\frac{g_{W}m_{W}}{\cos^{2}{\theta_{W}}} \;,
\;\;\;\;\;\; g_{W}=\frac{g_{e}}{\sin\theta_{W}}\;,
\;\;\;\;\;\;g_e=\sqrt{4\pi \alpha}
\end{eqnarray}
where $k_{1}^{\mu}$ and $k_{2}^{\mu}$ represent the momentum of two
W or Z boson (V=W, Z). For a convention we assume that all the
momenta are outgoing from the vertex. Within the standard model at
tree level, the couplings are given by $a_{V}=1$, $b_{V}=0$, and
$\beta_{V}=0$.

In the literature there have been several studies for anomalous {\it
WWH} couplings. The LHC will start operating soon. Anomalous gauge
couplings of the Higgs boson have been studied at the LHC via the
weak-boson scattering \cite{Yuan} and vector boson fusion
\cite{Hankele} processes. There has been a great amount of work on
anomalous {\it WWH} couplings which focus on the future linear
$e^{+}e^{-}$ collider and its $e\gamma$ and $\gamma\gamma$ modes.
Anomalous {\it WWH} couplings have been analyzed through the
processes $e^{+}e^{-} \to f\bar{f}H$ \cite{Choudhury1,Barger},
$e^{+}e^{-} \to W^{+} W^{-} \gamma$ \cite{Lietti}, $e^{-}\gamma \to
\nu_{e} W^{-} H$ \cite{Choudhury2}, and $\gamma \gamma \to WWWW$
\cite{Han}.

The anomalous cross sections are quadratic functions of the
anomalous parameters $\alpha_j\, (j=1,2,...)$ i.e.,

\begin{eqnarray}
\sigma_{AN}=\sigma_{SM}+\alpha_j\,\sigma^j_{int}+
\alpha_j^2\,\sigma^j_{ano}+\alpha_j\ \alpha_i
\,\sigma^{j\,i}_{ano}\nonumber
\end{eqnarray}
where $\sigma_{SM}$ is the SM cross section, $\sigma^j_{int}$ is the
interference term between the SM and the anomalous contribution, and
$\sigma^j_{ano}$ and $\sigma^{j\,i}_{ano}$ are the pure anomalous
contributions. In Ref. \cite{Choudhury2} authors ignore quadratic
anomalous coupling terms in the cross section calculations; they
retain contributions only up to the lowest nontrivial order, i.e.
$\alpha_j\,\sigma^j_{int}$. This is reasonable due to the higher
dimensional nature of their origin. On the other hand, these
quadratic terms generally have a higher momentum dependence than
linear terms and may have a significant contribution at high
energies. For example, in the anomalous {\it WWH} vertex (1)
coefficients of $\frac{b_{W}}{m_{W}^2}$ and
$\frac{\beta_{W}}{m_{W}^2}$ have a momentum dependence of dimension
2. Therefore in the anomalous cross section for the process
$e^{-}\gamma \to \nu_{e} W^{-} H$, the contributions
$\sigma^j_{ano}$ contain two additional momentum dependence when we
compare with $\sigma^j_{int}$. Of course the terms $\sigma^j_{ano}$
are suppressed by the inverse powers of new physics energy scale
(NPS) but their contribution  grows faster than the contribution of
$\sigma^j_{int}$ as the energy increases and approaches NPS.
Therefore these quadratic anomalous contributions should be
considered at high energy processes. We will consider all anomalous
terms in our tree-level cross section calculations.

As in Ref. \cite{Choudhury2}, we take into account initial electron
and photon polarizations before Compton backscattering. Furthermore
we take into account initial electron beam polarization which takes
part in the subprocess and also the final state polarizations of
{\it W} boson. We will show that these polarization configurations
which have not been considered in Ref. \cite{Choudhury2} lead to a
significant amount of improvement in the sensitivity limits.

\section{Polarized cross sections}

The process $e^{-}\gamma \to \nu_{e} W^{-} H$ takes part as a
subprocess in the $e^{+}e^{-}$ collision. A real gamma beam which
enters the subprocess is obtained through Compton backscattering of
laser light off linear electron or positron beam.

The spectrum of backscattered photons in connection with helicities
of initial laser photon and electron is given by \cite{Ginzburg}

\begin{eqnarray}
f_{\gamma/e}(y)={{1}\over{g(\zeta)}}[1-y+{{1}\over{1-y}}
-{{4y}\over{\zeta(1-y)}}+{{4y^{2}}\over {\zeta^{2}(1-y)^{2}}}+
\lambda_{0}\lambda_{e} r\zeta (1-2r)(2-y)]
\end{eqnarray}

where

\begin{eqnarray}
g(\zeta)=&&g_{1}(\zeta)+
\lambda_{0}\lambda_{e}g_{2}(\zeta) \nonumber\\
g_{1}(\zeta)=&&(1-{{4}\over{\zeta}}
-{{8}\over{\zeta^{2}}})\ln{(\zeta+1)}
+{{1}\over{2}}+{{8}\over{\zeta}}-{{1}\over{2(\zeta+1)^{2}}} \\
g_{2}(\zeta)=&&(1+{{2}\over{\zeta}})\ln{(\zeta+1)}
-{{5}\over{2}}+{{1}\over{\zeta+1}}-{{1}\over{2(\zeta+1)^{2}}}
\end{eqnarray}

Here $r=y/[\zeta(1-y)]$ and $\zeta=4E_{e}E_{0}/M_{e}^{2}$. $E_{0}$
and $\lambda_{0}$ are the energy and the helicity of the initial
laser photon and $E_{e}$ and $\lambda_{e}$ are the energy and the
helicity of the initial electron beam before Compton backscattering.
$y$ is the fraction which represents the ratio between the scattered
photon and initial electron energy for the backscattered photons
moving along the initial electron direction. The maximum value of
$y$ reaches 0.83 when $\zeta=4.8$ in which the backscattered photon
energy is maximized without spoiling the luminosity.

Backscattered photons are not in fixed helicity states. Their
helicities are described by a distribution :

\begin{eqnarray}
\xi(E_{\gamma},\lambda_{0})={{\lambda_{0}(1-2r)
(1-y+1/(1-y))+\lambda_{e} r\zeta[1+(1-y)(1-2r)^{2}]}
\over{1-y+1/(1-y)-4r(1-r)-\lambda_{e}\lambda_{0}r\zeta (2r-1)(2-y)}}
\end{eqnarray}

The helicity dependent differential cross section for the subprocess
can be connected to initial laser photon helicity $\lambda_{0}$ and
initial electron beam polarization $P_{e}$ through the formula

\begin{eqnarray}
&&d\hat{\sigma}(\lambda_{0},P_{e};\lambda_{W})\nonumber
\\ &&=\frac{1}{4}(1-P_{e})\left[(1+\xi(E_{\gamma},\lambda_{0}))
d\hat{\sigma}(+,L;\lambda_{W})+(1-\xi(E_{\gamma},\lambda_{0}))
d\hat{\sigma}(-,L;\lambda_{W})\right]\nonumber \\
&&+\frac{1}{4}(1+P_{e})\left[(1+\xi(E_{\gamma},\lambda_{0}))
d\hat{\sigma}(+,R;\lambda_{W})+(1-\xi(E_{\gamma},\lambda_{0}))
d\hat{\sigma}(-,R;\lambda_{W})\right]
\end{eqnarray}

Here $d\hat{\sigma}(\lambda_{\gamma},\sigma;\lambda_{W})$ is the
helicity dependent differential cross section in the helicity
eigenstates; $\sigma: L,R$, $\lambda_{\gamma}=+,-$ and
$\lambda_{W}=+,-,0$. It should be noted that $P_{e}$ and
$\lambda_{e}$ refer to different beams. $P_{e}$ is the electron beam
polarization which enters the subprocess but $\lambda_{e}$ is the
polarization of initial electron beam before Compton backscattering.

The integrated cross section over the backscattered photon spectrum
is

\begin{eqnarray}
d\sigma(\lambda_{0},P_{e};\lambda_{W})=\int_{y_{min}}^{0.83}
f_{\gamma/e}(y)d\hat{\sigma}(\lambda_{0},P_{e};\lambda_{W}) dy
\end{eqnarray}
where $y_{min}=\frac{(m_{W}+m_{H})^{2}}{s}$.

\begin{figure}
\includegraphics{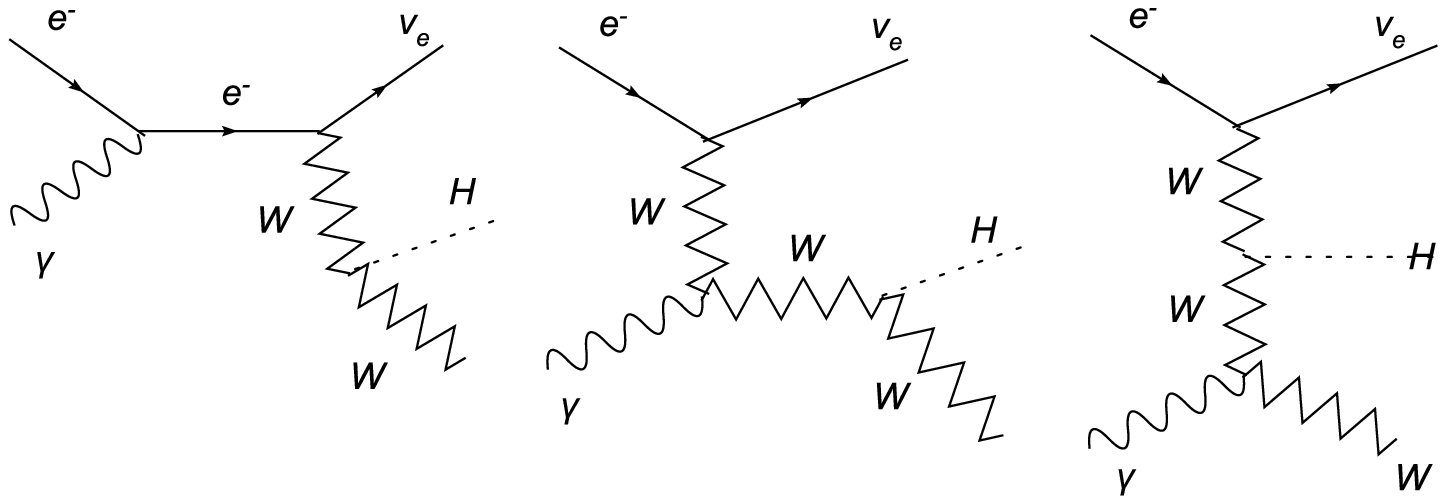}
\caption{Tree-level Feynman diagrams for $e^{-}\gamma \to \nu_{e}
W^{-} H$. \label{fig1}}
\end{figure}

The process $e^{-}\gamma \to \nu_{e} W^{-} H$ is described by three
tree-level diagrams (Fig.\ref{fig1}). Each of the diagrams contains
an anomalous {\it WWH} vertex. The helicity amplitudes have been
calculated using vertex amplitude techniques derived in Ref.
\cite{maina} and the phase space integrations have been performed by
{\sc GRACE} \cite{grace} which uses a Monte Carlo routine.

\begin{figure}
\includegraphics{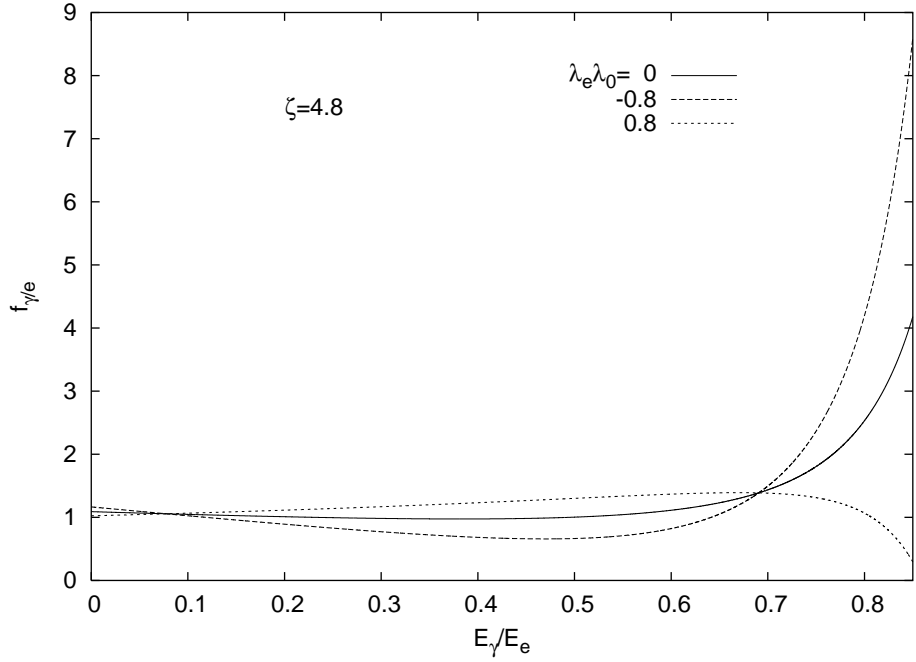}
\caption{Energy distribution of backscattered photons for
$\lambda_{e}\lambda_{0}=0, -0.8, 0.8.$ \label{fig2}}
\end{figure}

In our calculations we choose to work with a Higgs boson of mass 120
GeV which is compatible with current mass bounds. We accept that
initial electron beam polarizability is $|\lambda_{e}|,
|P_{e}|$=0.8. To see the influence of initial beam polarization,
energy distribution of backscattered photons $f_{\gamma /e}$ is
plotted for $\lambda_{e}\lambda_{0}$=0, -0.8, and +0.8 in
Fig.\ref{fig2}. We see from the figure that backscattered photon
distribution is very low at high energies in
$\lambda_{e}\lambda_{0}$=+0.8. Therefore we will only consider the
case $\lambda_{e}\lambda_{0}< 0$ in the cross section calculations.

One can see from Figs.\ref{fig3}-\ref{fig5} the influence of the
initial state polarizations on the deviations of the total cross
sections from their SM value. In Fig.\ref{fig3} initial polarization
configurations $(\lambda_{e},\lambda_{0},P_{e})=(+0.8,-1,\mp0.8)$
are omitted since they coincide with
$(\lambda_{e},\lambda_{0},P_{e})=(-0.8,+1,\mp0.8)$. We see from
Figs. \ref{fig3}-\ref{fig5} that cross section is very sensitive to
$P_{e}$. This is reasonable due to $V-A$ structure of the
$We\nu_{e}$ vertex in the Feynman diagrams. Deviation of the cross
section from its SM value reaches its maximum at the
$(\lambda_{e},\lambda_{0},P_{e})=(-0.8,+1,-0.8)$ polarization
configuration. In Figs.\ref{fig6}-\ref{fig8} we plot the total cross
section as a function of anomalous parameters for various final
state polarizations. In these figures TR and LO stand for
"transverse" and "longitudinal," respectively. We see from these
figures that cross section is very small in the longitudinal
polarization configuration. On the other hand, longitudinal
polarization remarkably improves the deviations from the SM at the
positive values of the parameter $b_{W}$. For instance, in
Fig.\ref{fig7} longitudinal cross section increases by a factor of
12 as $b_{W}$ increases from 0 to 0.21. But this increment is only a
factor of 1.4 for the unpolarized case. The reason for this comes
from the fact that the longitudinal cross section has a symmetric
behavior in the positive and negative intervals of the parameter but
the unpolarized graph shifts a little to the right. Longitudinal
polarization also improves the deviations from the SM at both
positive and negative values of $\beta_{W}$.

Angular distributions of the Higgs boson for unpolarized and
longitudinally polarized cross sections are given in Figs.\ref{fig9}
and \ref{fig10}. In the figures $\theta_{H}$ is the angle between
the outgoing Higgs boson and the incoming electron in the center of
mass frame of $e^{+}e^{-}$. Angular distributions for other
polarization configurations are very similar to the unpolarized
cross section. So we will not give them in the paper. One can see
from Fig.\ref{fig10} that deviation of the differential cross
section from its SM value is larger in the negative interval of
$\cos\theta_{H}$.

\section{Limits on the anomalous coupling parameters}

For a concrete result we have obtained 95\% C.L. limits on the
anomalous coupling parameters $a_{W}$, $b_{W}$, and $\beta_{W}$
using $\chi^{2}$ analysis at $\sqrt{s}=0.5$ and 1 TeV and integrated
luminosity $L_{int}=500$ $fb^{-1}$ without systematic errors. In our
calculations we choose to work with a Higgs boson of mass 120 GeV.
Therefore the dominant decay mode should be $H\to b\bar{b}$ with a
branching ratio $B_{H}\approx0.9$.

We assume that W polarization can be measured. Indeed angular
distribution of the W decay products has a clear correlation with
the helicity states of the W boson. Therefore we consider the case
in which W momentum is reconstructible. We restrict ourselves to a
$W\to q\bar{q}^\prime$ decay channel with a branching ratio
$B_{W}\approx0.68$. The number of events are given by $N=E
B_{W}B_{H}L_{int}\sigma$, where $E$ is the {\it b}-tagging
efficiency and it is taken to be 0.7 as Refs.
\cite{Choudhury1,Choudhury2}. There have been several experimental
studies in the literature for the measurement of W polarization. W
boson polarization has been studied at the CERN $e^+ e^-$ collider
LEP2 via the process $e^+ e^- \to W^+ W^- \to \ell \nu q \bar
{q}^\prime$ \cite{LEP}. At Fermilab Tevatron, polarization of the W
bosons produced in the top quark decay has been measured by the CDF
and D{\O} collaborations \cite{Tevatron}.

\begin{figure}
\includegraphics{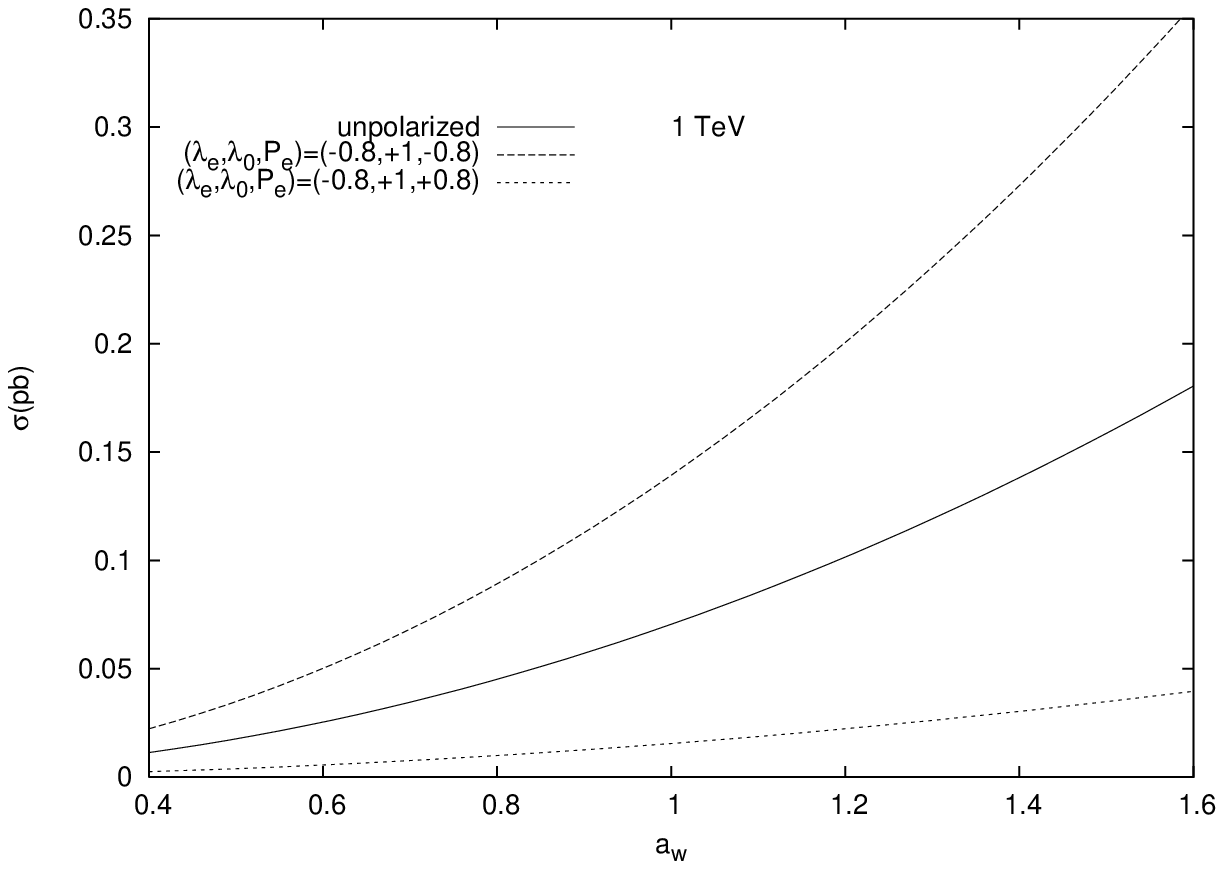}
\caption{The integrated total cross section of $e^{-}\gamma \to
\nu_{e} W^{-} H$ as a function of anomalous coupling $a_{W}$. The
legends are for initial beam polarizations. $\sqrt{s}= 1 TeV$.
\label{fig3}}
\end{figure}

\begin{figure}
\includegraphics{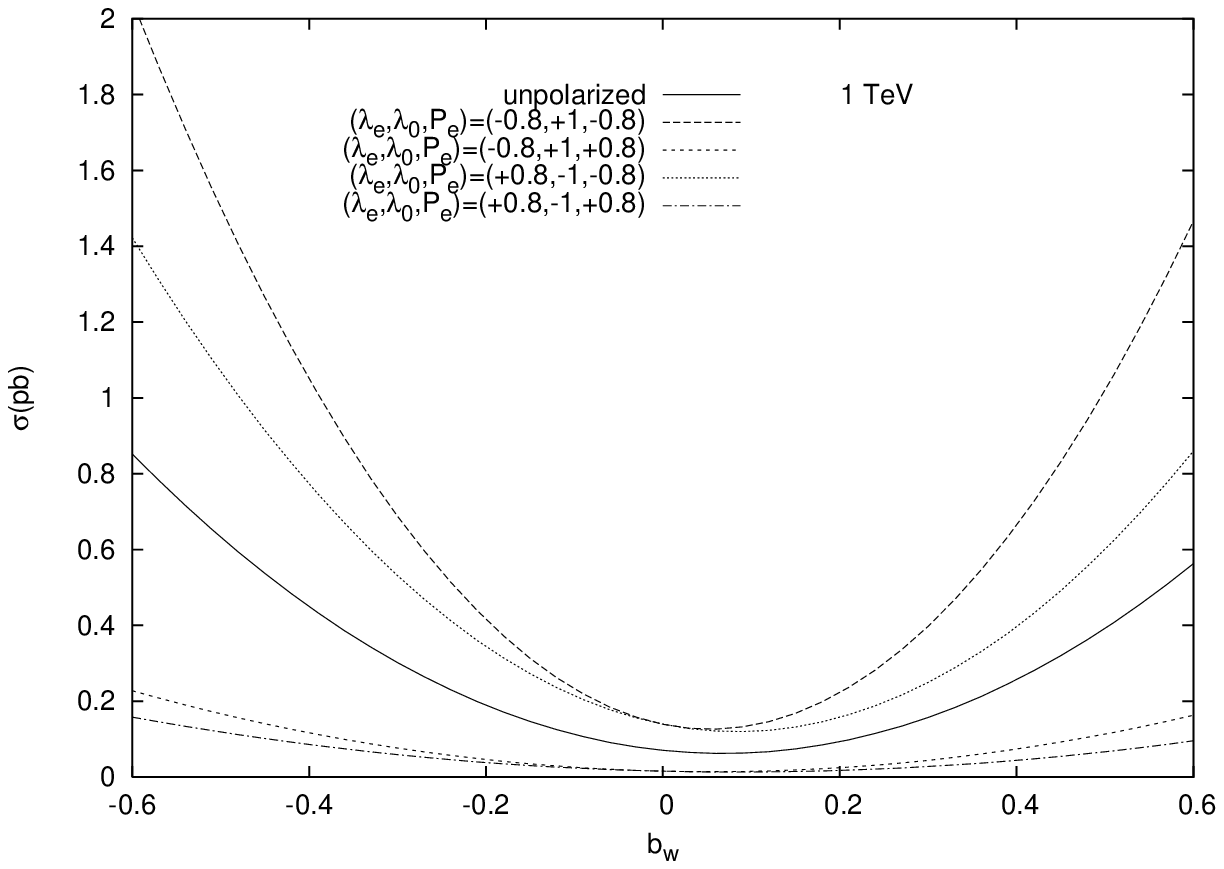}
\caption{The integrated total cross section of $e^{-}\gamma \to
\nu_{e} W^{-} H$ as a function of anomalous coupling $b_{W}$. The
legends are for initial beam polarizations. $\sqrt{s}= 1 TeV$.
\label{fig4}}
\end{figure}

\begin{figure}
\includegraphics{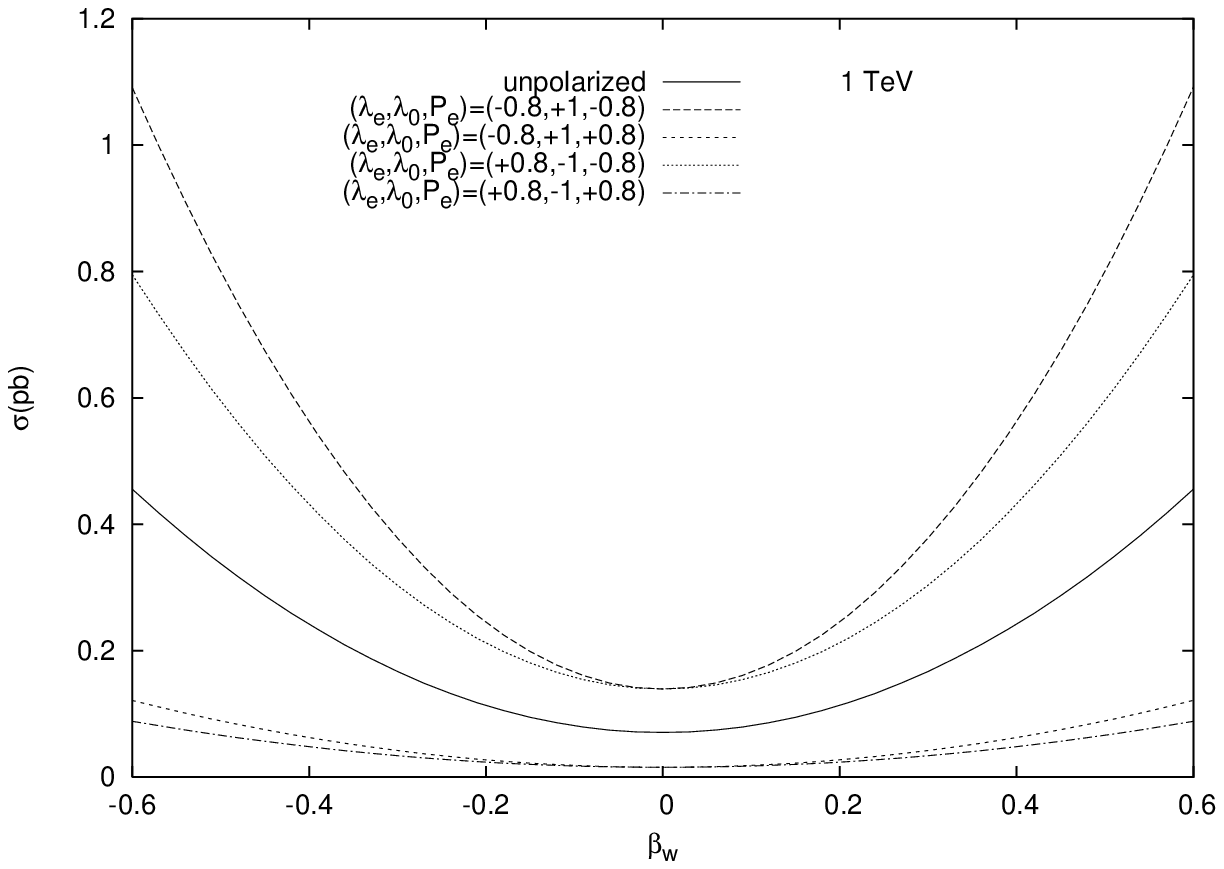}
\caption{The integrated total cross section of $e^{-}\gamma \to
\nu_{e} W^{-} H$ as a function of anomalous coupling $\beta_{W}$.
The legends are for initial beam polarizations. $\sqrt{s}= 1 TeV$.
\label{fig5}}
\end{figure}

\begin{figure}
\includegraphics{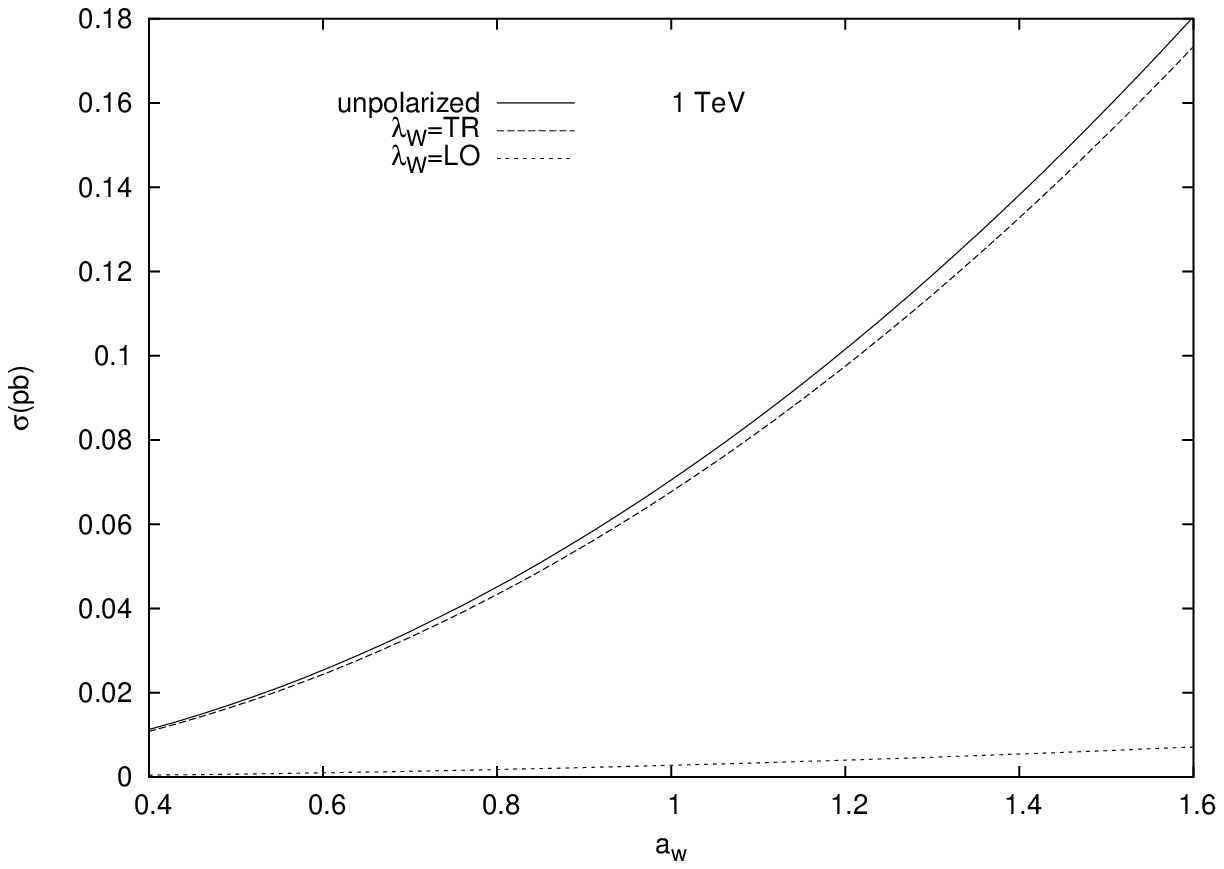}
\caption{The integrated total cross section of $e^{-}\gamma \to
\nu_{e} W^{-} H$ as a function of anomalous coupling $a_{W}$. The
legends are for final state polarizations. $\sqrt{s}= 1
TeV$.\label{fig6}}
\end{figure}

Limits on the anomalous {\it WWH} couplings are given in Tables
\ref{tab1} and \ref{tab2} for various initial and final polarization
configurations. In the tables,
$(\lambda_{0},\lambda_{e},P_{e})=(0,0,0)$ is for unpolarized initial
beams and TR+LO represents the unpolarized W boson. We take into
account  $(\lambda_{0},\lambda_{e},P_{e})=(+1,-0.8,-0.8)$ initial
polarization configuration which gives the largest deviation from
the SM (Figs. \ref{fig3}-\ref{fig5}). It can be seen from
Fig.\ref{fig10} that deviation of the differential cross section
from its SM value is larger in the negative interval of
$\cos\theta_{H}$. So we impose a restriction $\cos\theta_{H}<0$ on
the longitudinal polarization configuration to improve the limits.
We represent these restricted limits with a superscript " * ". We
see from Table \ref{tab1} that polarization leads to a significant
improvement on the upper bound of $b_{W}$. Polarization improves the
upper bound of $b_{W}$ by a factor of 7.4. This improvement factor
becomes 9.3 when we consider $\cos\theta_{H}<0$ restriction. At
$\sqrt{s}=0.5$ TeV these improvement factors are smaller; we see
from Table \ref{tab2} that initial state polarization together with
final polarization and $\cos\theta_H < 0$ restriction improve the
upper bound of $b_W$ approximately by a factor of 4.3. Polarization
improves both upper and lower bounds of $\beta_{W}$. Improvement
factors for upper and lower bounds are the same and equal to 2.2 at
$\sqrt{s}=1$ TeV and 1.9 at $\sqrt{s}=0.5$ TeV. Final state
polarizations do not improve the lower bound of $b_{W}$. But initial
polarizations improve the lower bound of $b_{W}$ by a factor of 1.7
at $\sqrt{s}=1$ TeV and 1.6 at $\sqrt{s}=0.5$ TeV.

\begin{figure}
\includegraphics{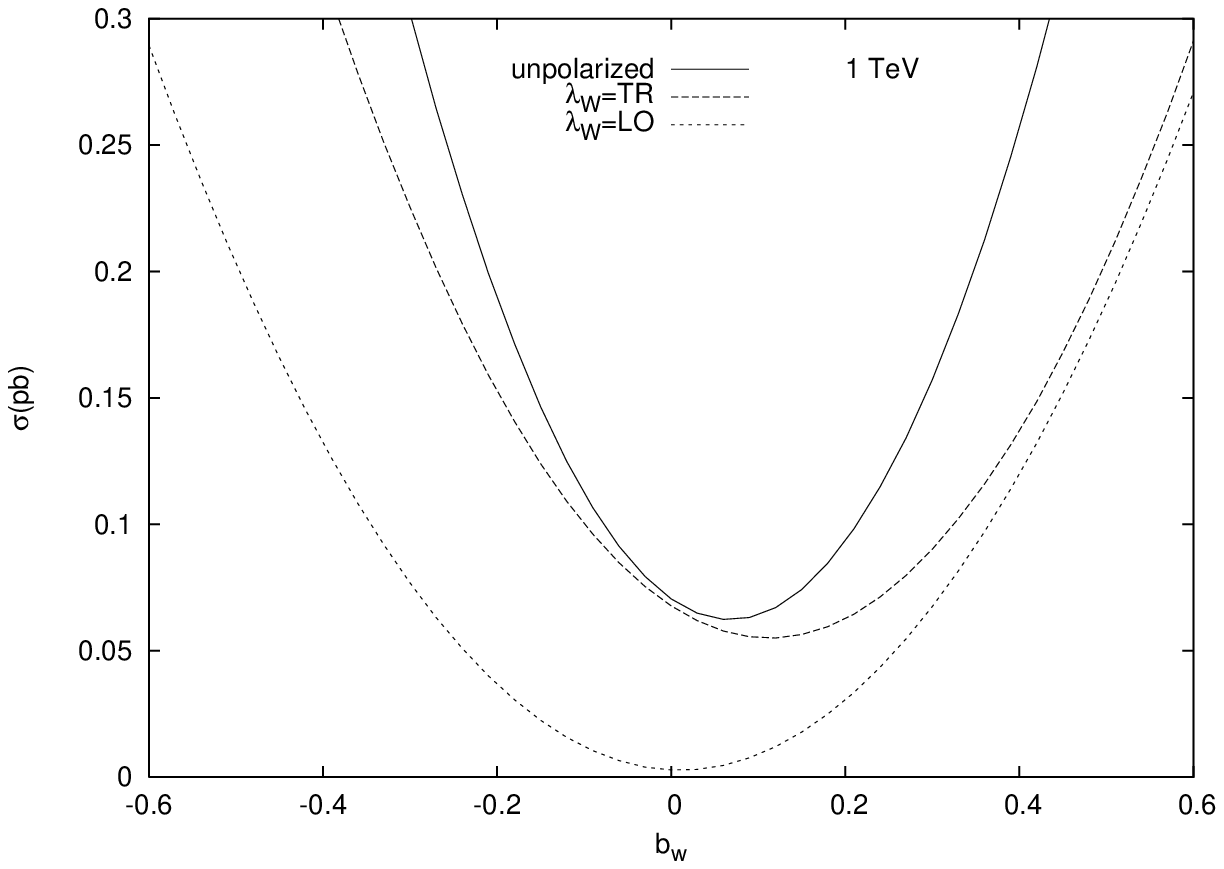}
\caption{The integrated total cross section of $e^{-}\gamma \to
\nu_{e} W^{-} H$ as a function of anomalous coupling $b_{W}$. The
legends are for final state polarizations. $\sqrt{s}= 1 TeV$.
\label{fig7}}
\end{figure}

\begin{figure}
\includegraphics{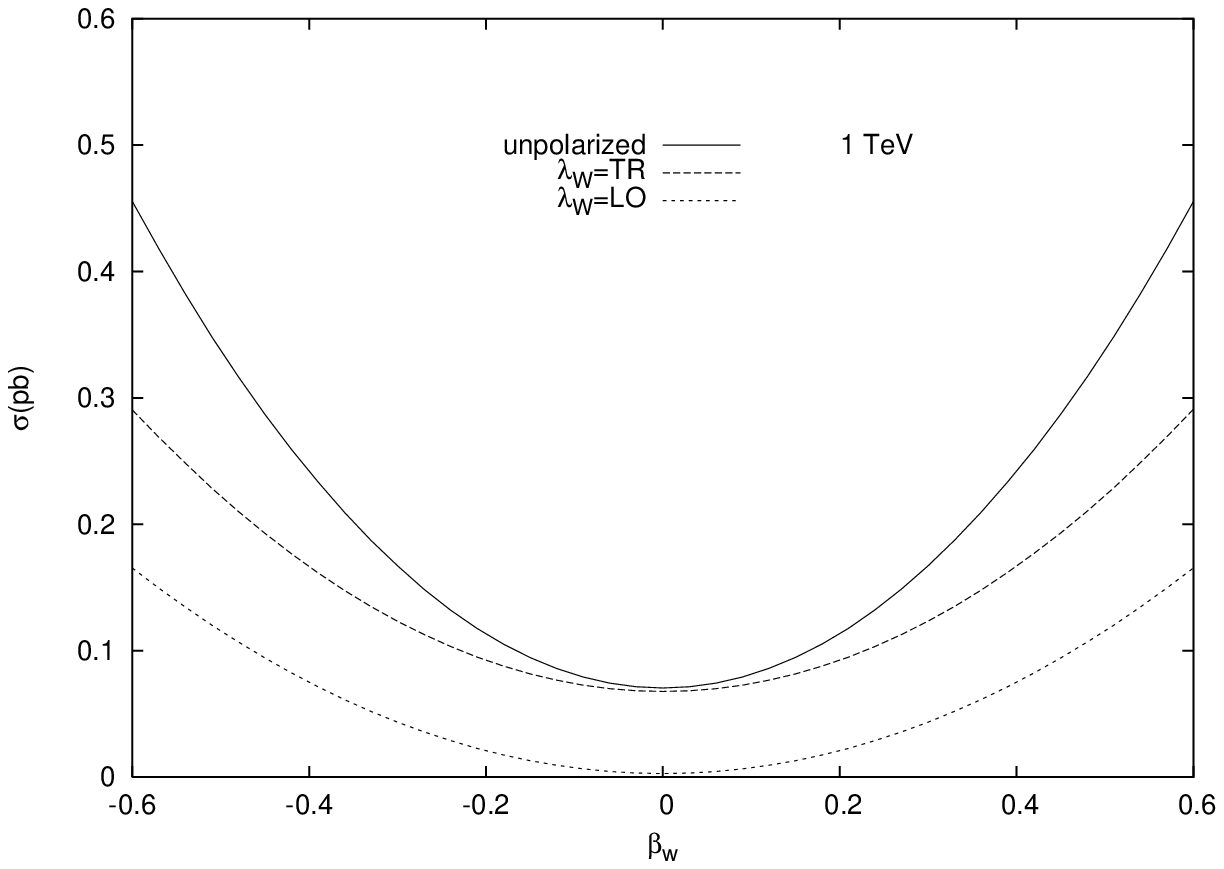}
\caption{The integrated total cross section of $e^{-}\gamma \to
\nu_{e} W^{-} H$ as a function of anomalous coupling $\beta_{W}$.
The legends are for final state polarizations. $\sqrt{s}= 1 TeV$.
\label{fig8}}
\end{figure}

\begin{figure}
\includegraphics{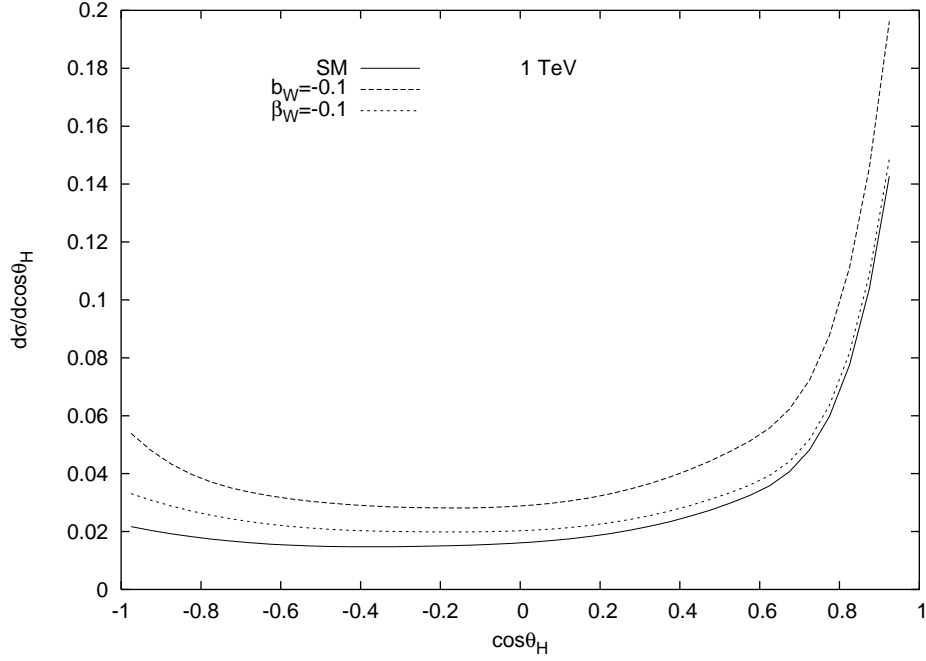}
\caption{Angular distributions of the Higgs boson in the center of
the mass frame of $e^{+}e^{-}$ for unpolarized beams. $\theta_{H}$
is the angle between the outgoing Higgs boson and the incoming
electron. \label{fig9}}
\end{figure}

\begin{figure}
\includegraphics{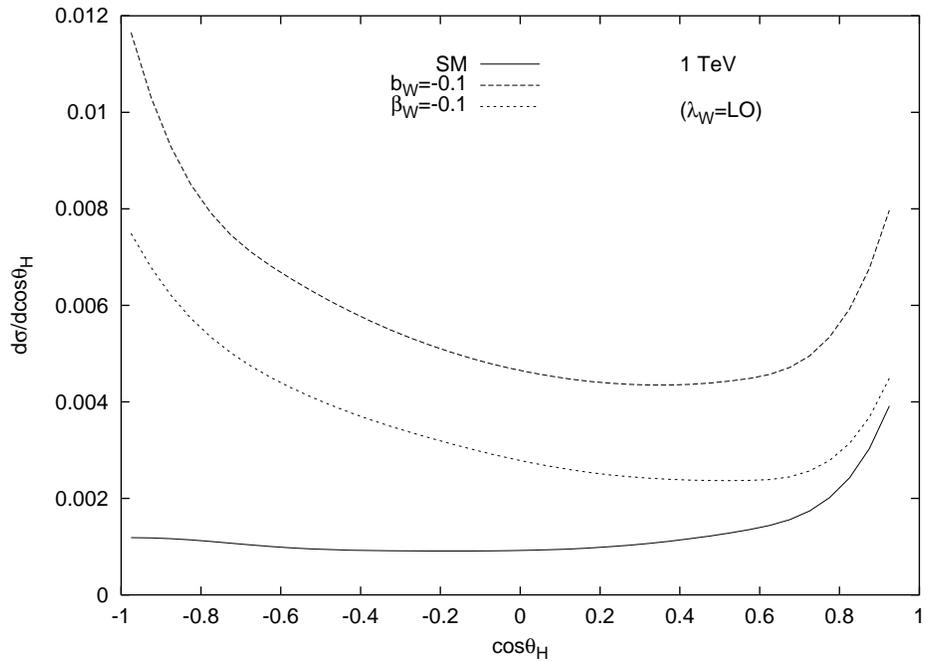}
\caption{Angular distributions of the Higgs boson in the center of
mass frame of $e^{+}e^{-}$ for the longitudinal polarization state
of the final W. \label{fig10}}
\end{figure}

Anomalous  couplings were studied in \cite{Choudhury2} through the
same process $e^{-}\gamma \to \nu_{e} W^{-} H$ at the linear order
of anomalous couplings and $\sqrt{s}=0.5$ TeV energy. Authors took
into account initial electron and photon polarizations ($\lambda_e$
and $\lambda_0$) before Compton backscattering but they neglected
initial electron beam polarization ($P_e$) and also the final state
polarization. Imposing same acceptance cuts, we have confirmed the
results of \cite{Choudhury2} for all combinations of the initial
polarizations $\lambda_e$ and $\lambda_0$. The unpolarized cross
section at the linear order and $\sqrt{s}=0.5$ TeV energy is given
by

\begin{eqnarray}
\sigma=[2.86(1+2\Delta a_W)-11.07\,b_W]fb
\end{eqnarray}
where $\Delta a_W =a_W -1$ and coefficient of the parameter
$\beta_W$ is zero at the linear order due to $\hat{T}$ invariance of
the total cross section. $\hat{T}$ is the pseudo-time reversal
transformation, one which reverses particle momenta and spins but
does not interchange initial and final states. The term proportional
to $\beta_W$ is $\hat{T}$ odd in the total cross section at the
linear order. Therefore only the quadratic order terms for $\beta_W$
contribute to the cross section. In \cite{Choudhury2} authors
imposed some cuts and restrict themselves to an appropriate part of
the phase space in order to see the effects of $\beta_W$. It is,
however, irrelevant for our treatment since we have considered all
anomalous terms in our calculations. Using cross section (9) authors
set $3\sigma$ bounds on the anomalous coupling parameters $\Delta
a_W$ and $b_W$ with an integrated luminosity of 500 $fb^{-1}$. If we
convert these bounds into 95\% confidence level they are
approximately $|\Delta a_W|<0.026$ and $|b_W|<0.013$. We see from
Table \ref{tab2} that the limit on $\Delta a_W$ is very close to the
limit at the linear order and quadratic order anomalous contribution
leads to a slight improvement. On the other hand, different from the
limits at the linear order, limits on $b_W$ are not symmetric. The
lower bound of $b_{W}$ is improved slightly but the upper bound get
worse markedly. This is very compatible with Figs. \ref{fig4} and
\ref{fig7} where we observe that minima of the unpolarized graphs
shift a little to the right.

In conclusion, polarization leads to a considerable improvement on
the upper bound of the anomalous parameter $b_{W}$. It improves both
upper and lower bounds of $\beta_{W}$. Although the SM cross
sections in the longitudinal polarization configuration of the final
W are small, sensitivity limits are better than the transverse
polarization case. As mentioned in \cite{Choudhury1,Barger} the
process $e^- e^+ \to \nu \bar{\nu} H$ can be a good probe to measure
the {\it WWH} coupling at high energies. The disadvantage of $e^-
e^+ \to \nu \bar{\nu} H$ is that it also receive contributions from
the {\it ZZH} vertex and with the final neutrinos being invisible
this process has too few observables associated with it. This makes
it very difficult to dissociate {\it WWH} from {\it ZZH}
\cite{Choudhury1}.


\begin{table}
\caption{Sensitivity of the  $e\gamma $ collision to anomalous $WWH$
couplings at 95\% C.L. for $\sqrt{s}=1$ TeV and $L_{int}=500$
$fb^{-1}$. The effects of final state W boson polarization and
initial beam polarizations are shown in each row. The superscript "
* " represents the restriction $\cos\theta_{H}<0$ in the phase
space. \label{tab1}}
\begin{ruledtabular}
\begin{tabular}{ccccccc}
$\lambda_{0}$&$\lambda_{e}$ & $P_{e} $& $\lambda_{W}$& $a_{W}$
& $b_{W}$& $\beta_{W}$\\
\hline
  0&0  &0  &$TR+LO$  &(0.992, 1.008)  &(-0.005, 0.140) &(-0.033, 0.033)  \\
  0&0  &0  &$TR$  &(0.992, 1.008)  &(-0.005, 0.230) &(-0.043, 0.043)  \\
  0&0  &0  &$LO$  &(0.960, 1.040)  &(-0.010, 0.030) &(-0.023, 0.023)  \\
  0&0  &0  & $LO^{*}$  &$(0.931,1.065)^{*}$  &$(-0.011,0.022)^{*}$ &$(-0.020,0.020)^{*}$  \\
  +1&-0.8 &-0.8  &$TR+LO$  &(0.995, 1.006)  &(-0.003, 0.111) &(-0.024, 0.024)  \\
  +1&-0.8 &-0.8  &$TR$  &(0.995, 1.006)  &(-0.003, 0.205) &(-0.033, 0.033)  \\
  +1&-0.8 &-0.8  &$LO$  &(0.970, 1.030)   &(-0.007, 0.019) &(-0.015, 0.015)  \\
  +1&-0.8 &-0.8  &$LO^{*}$  &$(0.953, 1.047)^{*}$   &$(-0.008, 0.015)^{*}$ &$(-0.014, 0.014)^{*}$  \\
\end{tabular}
\end{ruledtabular}
\end{table}

\begin{table}
\caption{The same as Table I but for $\sqrt{s}=0.5$ TeV.
\label{tab2}}
\begin{ruledtabular}
\begin{tabular}{ccccccc}
$\lambda_{0}$&$\lambda_{e}$ & $P_{e} $& $\lambda_{W}$& $a_{W}$
& $b_{W}$& $\beta_{W}$\\
\hline
  0&0  &0  &$TR+LO$  &(0.980, 1.020)  &(-0.011, 0.261) &(-0.078, 0.078)  \\
  0&0  &0  &$TR$  &(0.980, 1.020)  &(-0.012, 0.363) &(-0.098, 0.098)  \\
  0&0  &0  &$LO$  &(0.930, 1.067)  &(-0.018, 0.114) &(-0.067, 0.067)  \\
  0&0  &0  & $LO^{*}$  &$(0.865, 1.120)^{*}$  &$(-0.020, 0.087)^{*}$ &$(-0.060, 0.060)^{*}$  \\
  +1&-0.8 &-0.8  &$TR+LO$  &(0.987, 1.014)  &(-0.007, 0.206) &(-0.056, 0.056)  \\
  +1&-0.8 &-0.8  &$TR$  &(0.986, 1.013)  &(-0.008, 0.307) &(-0.085, 0.085)  \\
  +1&-0.8 &-0.8  &$LO$  &(0.950, 1.050)   &( -0.011, 0.077) &(-0.042, 0.042)  \\
  +1&-0.8 &-0.8  &$LO^{*}$  &$(0.910, 1.070)^{*}$   &$(-0.012, 0.061)^{*}$ &$(-0.058, 0.058)^{*}$  \\
\end{tabular}
\end{ruledtabular}
\end{table}

\end{document}